# Role of Autoconversion Parameterization in Coupled Climate Model for Simulating Monsoon Subseasonal Oscillations

Ushnanshu Dutta[1*], Moumita Bhowmik[2], Anupam Hazra[1,2*], Suryachandra. A. Rao[2], and Jen-Ping Chen[1,3,4]

[1]Department of Atmospheric Sciences, National Taiwan University, Taipei, Taiwan.

[2]Indian Institute of Tropical Meteorology, Ministry of Earth Sciences, Pune, India.

[3] Research Center for Environmental Changes, Academia Sinica, Taiwan

[4] International Degree Program in Climate Change and Sustainable Development, National Taiwan University, Taiwan

Corresponding authors:

Dr. Ushnanshu Dutta (duttaushnanshu@ntu.edu.tw, ushnanshudutta@gmail.com)

Dr. Anupam Hazra (hazra@tropmet.res.in)

**Key Points**

1. The proper combination of autoconversion coefficients improves the simulation of subseasonal oscillations of the Indian summer monsoon (ISM).

2. The Liu-Daum autoconversion rate provides better results in simulating the MISO characteristics in a high-resolution setup of CFSv2.

3. The results demonstrate a pathway for improving the simulation of subseasonal oscillations of ISM.

**Abstract**

The Indian summer monsoon (ISM) and associated monsoon intraseasonal oscillations (MISOs) influence the billions of people living in the Indian subcontinent. This study explores the role of autoconversion parameterization in microphysical schemes for the simulation of MISO with the coupled climate model, e.g., the Climate Forecast System version 2 (CFSv2), by conducting sensitivity experiments in two resolutions (~100 km and ~38 km). Results reveal that the modified autoconversion parameterization better simulates the active-break spells of the ISM rainfall. The main improvements include the contrasting features of rainfall over land and ocean and the MISO index, representing MISO periodicity. The improvements are qualitatively and quantitatively more significant in the higher-resolution simulations, particularly regarding rainfall spatial patterns over the Indian subcontinent during active spells. The MISO monitoring index in the revised CFSv2 also shows improvement compared to the control run. This study concludes that proper autoconversion parameterization in the coupled climate model can lead to enhanced representation of active/break spells and sub-seasonal variability of ISM.

**Plain Language Summary:**

Prediction of the Indian summer monsoon (ISM) and associated subseasonal variabilities is very important to policymakers and common people. For better simulation, cloud microphysical processes associated with rainfall must be parameterized properly. Autoconversion is a crucial cloud microphysical process that controls rain formation. In this paper, we have shown the importance of the proper combination of convective and microphysical autoconversion coefficients for better simulation of subseasonal oscillation of ISM in a coupled climate model, e.g., CFSv2.

We have also compared two different types of autoconversion parameterization in terms of simulating the dry and wet spells in a higher-resolution version of CFSv2. Results show the proper choice of autoconversion parametrization is crucial for improving the simulation of ISM.

# 1. Introduction

The prediction of the active-break spell of the Indian summer monsoon (ISM) is highly beneficial to the agricultural sector of the Indian subcontinent for proper planning of crop harvesting, and policymakers are also interested in the seasonal mean ISM. Reliable seasonal forecasting also depends on the extent of realistic simulation of active break spells, which determines the seasonal mean ISM. Over the past few decades, the prediction of the Indian Summer Monsoon (ISM) has been gradually improved (Rajeevan et al., 2012; Rao et al., 2019). Numerical weather prediction has also evolved in the last fifty years because of a billion-fold increase in computation power and the availability of high-performance computers (HPCs) (Michalakes, 2020). The scientific communities across the world are trying to improve model physics and parameterization schemes in climate models (Eyring et al., 2016; Meehl et al., 2007; Taylor et al., 2012). The incorporation of higher spatial resolution and new physical processes in the models participating in the Coupled Model Intercomparison Project (CMIP) has shown steady improvements in simulating the Asian monsoon with updated versions (Dutta et al., 2022; Gusain et al., 2020; Sperber et al., 2013; Zhu & Yang, 2021). (Dutta et al., 2024) also shown that improved mean characteristics in some latest generation CMIP6 models is due to better representation of MISO. However, (S. K.Saha et al., 2019)) pointed out that the latest generation models still need help to predict even 70% of the interannual variability of ISM rainfall (ISMR).

Studies (Hazra, Chaudhari, Saha, &Pokhrel, 2017; Hazra, Chaudhari, Saha, Pokhrel, et al., 2017; Maloney &Hartmann, 2001) have found that the poor representation of the cloud process inside a parameterization scheme in a coupled global climate model (CGCM) is one of the major sources of these limitations. The importance of proper representation of the autoconversion process in the model parameterization scheme has been suggested in this regard (Boyle et al., 2015; Ganai et al., 2019; J. Y. Han et al., 2016; S. Y. Hong et al., 2018; Michibata & Takemura, 2015; Rotstayn, 2000; Weber & Quaas, 2012; Zhang et al., 2002). (Song &Zhang, 2011) also indicated that the feedback between the microphysical parameterization scheme and the convective parameterization scheme is crucial to realistically simulating the convective to stratiform rain ratio. Therefore, (Dutta et al., 2021) designed sensitivity experiments based on a different combination of convective and microphysical "autoconversion" in the CFSv2 to improve the biases in the simulation of tropical oscillations (i.e., MISO and MJO) in the standard CFSv2 model. They identified a combination of convective autoconversion coefficient ($C_{CA}$) and microphysical autoconversion coefficient ($C_{MA}$), which are more realistic for simulating ISM mean features. However, it was unraveled how the modified set of autoconversion coefficients can impact the active-break spell of ISM. Therefore, this study investigates the impact of modified autoconversion rates (Sundqvist et al., 1989) on the simulation of active-break spells of ISM.

(Ramu et al., 2016) showed that increasing the horizontal resolution in CFSv2 can produce better simulation and prediction of the ISMR. On the other hand, (Liu et al., 2006) generalized the autoconversion rate, which is dependent on cloud water content, droplet number concentration, and relative dispersion of cloud droplets, where the value of relative dispersion to formulate Liu and Daum autoconversion rate can be obtained from Small-scale Lagrangian particle-based numerical simulation. (Liu et al., 2004, 2006) proposed that an autoconversion scheme based on relative dispersion can yield better results. (Bhowmik et al., 2024)) also demonstrated that the Sundqvist-type autoconversion rate fails to differentiate between shallow and convective clouds, whereas the Liu and Daum autoconversion can effectively represent these distinct cloud types, which can improve the mean ISM characteristics. Therefore, the scientific question arises: Can a high-resolution model with more generalized autoconversion parameterization be more useful in simulating the active break spell? To address it, we have replaced the cloud water rainwater autoconversion of (Sundqvist et al., 1989)) with a dispersion-based Liu-Damn autoconversion rate (Liu &Daum, 2004), and the sensitivity experiments are carried out in high-resolution (∼ 38 km) CFSv2 (T382). This provides the future direction of a high-resolution climate model with dispersion-based autoconversion parameterization (Liu et al., 2006; Liu &Daum, 2004) for the simulation of MISO. We believe the current study will provide new insight to the scientific community regarding the seasonal forecasting of ISM.

**2. Data and Methodology:**

For model performance evaluation, we focus on daily data of rainfall, high cloud fraction (HCF), outgoing longwave radiation (OLR), cloud mixing ratio (ice and water), and pressure vertical velocity (omega). Rainfall data are taken from the Global Precipitation Climatology Product (GPCP, (Adler et al., 2003). OLR data are from the National Oceanic and Atmospheric Administration interpolated data (Liebmann &Smith, 1996). HCF data are taken from the recently released fifth generation of the European Centre for Medium-Range Weather Forecasts (ECMWF) reanalysis, ERA5 (Hersbach et al., 2020). The daily data of the convective component of rainfall, total rainfall, cloud ice and water mixing ratio are also obtained from the ERA5. The daily data of rainfall from the Tropical Rainfall Measuring Mission (TRMM) are also considered in the study (Huffman et al., 2007). The observational/reanalysis is considered for ten years, i.e., from 1999 to 2008.

For the simulations, we applied the coupled climate model from the National Centers for Environmental Prediction (NCEP) Climate Forecast System version 2 (CFSv2) (S. Saha et al., 2014), which is comprised of a spectral atmospheric model with 64 hybrid vertical levels and the Geophysical Fluid Dynamics Laboratory (GFDL) Modular Ocean Model, version 4p0d (Griffies et al., 2005). The new Simplified Arakawa Schubert (NSAS, (J.Han &Pan, 2011) convective parameterization scheme in conjunction with mixed-phase cloud microphysics schemes (Zhao &Carr, 1997) within the CFSv2 model are considered here.

The design of sensitivity experiments using low-resolution (T126) CFSv2 simulation is based on different combinations of convective and microphysical autoconversion coefficients (Dutta et al., 2021). They designed six sensitivity experiments using a combination of three different values of $C_{CA}$ (viz., 0.0005 m$^{-1}$, 0.001 m$^{-1}$, 0.002m$^{-1}$) and three different values of $C_{MA}$ (viz., 1.0 x 10$^{-4}$ s$^{-1}$, 1.5 x 10$^{-4}$ s$^{-1}$, 2.5 x 10$^{-4}$ s$^{-1}$). Out of the six sensitivity experiments, the combination having $C_{CA}$= 0.001 m$^{-1}$ and $C_{MA}$= 1.5 x 10$^{-4}$ s$^{-1}$ performed the best in simulating the MISO and MJO characteristics along with improved mean simulation of the ISM. This combination, i.e., the modified physics version or CFSv2.MPHY also improves the teleconnection of cloud hydrometeors with global predictors, which is encouraging regarding the seasonal prediction of ISM (Hazra et al., 2023). Therefore, the combination of autoconversion coefficients for two sensitivity experiments (SE) with the CFSv2-T126 model (SE126 hereafter) in this current study is as follows:

a) CFSv2.CTL: $C_{CA}$= 0.002 m$^{-1}$; $C_{MA}$ = 1.0 x 10$^{-4}$ s$^{-1}$

b) CFSv2.MPHY: $C_{CA}$= 0.001 m$^{-1}$; $C_{MA}$ = 1.5 x 10$^{-4}$ s$^{-1}$.

The study also contains a preliminary analysis of another set of sensitivity experiments with high resolution (~ 38 km) CFSv2-T382 model (SE382 hereafter), which compares two different types of autoconversion schemes, i.e., (Sundqvist et al., 1989) and Liu-Daum scheme (Liu et al., 2006) simulating the active-break conditions. The details regarding the setup of the Liu-Daum autoconversion scheme in CFSv2-T382 can be obtained from the recent study by (Bhowmik et al., 2024). The CFSv2-T382 model with the Sundqvist et al. and Liu-Daum schemes is termed CFSv2-SQ-HR and CFSv2-LD-HR, respectively.

In each sensitivity experiment, the CFSv2 is initialized using the same initial conditions for the ocean and atmosphere from NCEP Climate Forecast System Reanalysis (CFSR,(S.Saha et al., 2010), and the model is integrated for 15 years. We have excluded the first five years for spin-up purposes. For all the simulations physical processes like land surface, radiation, etc. are kept unchanged.

Active and break events were calculated, as mentioned in the earlier study by Dutta et al., (2020). The average daily rainfall anomaly is averaged over the central Indian core monsoon region (74° E–86° E and 16° N–26° N) followed by division by the standard deviation. The active and break spells are considered as the periods during which the standardized daily rainfall (unfiltered) anomaly is more than + 1.0 and less than − 1.0, respectively, for consecutively three days or more (Rajeevan et al., 2010). Further, we have also analyzed the fidelity of the simulation of MISO indices as described by (Suhas et al., 2013). The extended empirical orthogonal function (EEOF) analysis has been applied to the ten-year unfiltered daily rainfall data for June-July-August-September (JJAS). The rainfall data was initially averaged over Indian longitudes (65˚E-95˚E). The principal components (PC) 1 and 2 are MISO1 and MISO2 indices. The details in this regard, i.e., PC selection, are available in the study by (Suhas et al., 2013).

**3. Results**

## 3.1. Active-Break Features

### 3.1.1. Rainfall and Convection

We first analyzed the representation of rainfall patterns during active (wet) and break (dry) spells from the observation and SE126. The active composite of rainfall anomaly from observation (GPCP) and SE126 are shown in Figure 1. The positive (negative) anomaly identifies anomalously more (less) rainfall. During active spells, most parts of the Indian landmass and the Bay of Bengal (BoB) receive anomalously high rainfall (Fig. 1a). High rainfall is also noticed over the Western Ghats (WG) and the adjacent Arabian Sea (AS). The equatorial Indian Ocean adjacent to maritime continents receives anomalously low rainfall during active spells. The observation also shows a negative rainfall anomaly over the southeastern part of the Tibetan Plateau (TP), which suggests a dipole pattern during the active spell of the ISMR (Jiang & Ting, 2019). The extent of positive rainfall anomaly is primarily confined only to the central India region in CSFv2.CTL. It also underestimates the high rainfall band near the WG. Contrastingly, it shows negative rainfall anomalies over the majority of the area of the BoB. The dipole structure of southeastern TP and central India is also not captured by CFSv2.CTL. CFSv2.MPHY (Fig. 1c) shows improvement in simulating the observed features in active spell than CFSv2.CTL. The spatial extent of positive rainfall over the Indian subcontinent is better in CFSv2.MPHY. Besides, it can also capture the dipole pattern of rainfall anomaly between southeastern TP and central India (Fig. 1c). The pattern correlation coefficient (PCC) with observation is also higher for CFSv2.MPHY (~ 0.8) than that of CFSv2.CTL (~ 0.6) over the Indian subcontinent box (Fig. 1b, 60˚E-100˚E, 10˚S-30˚N).

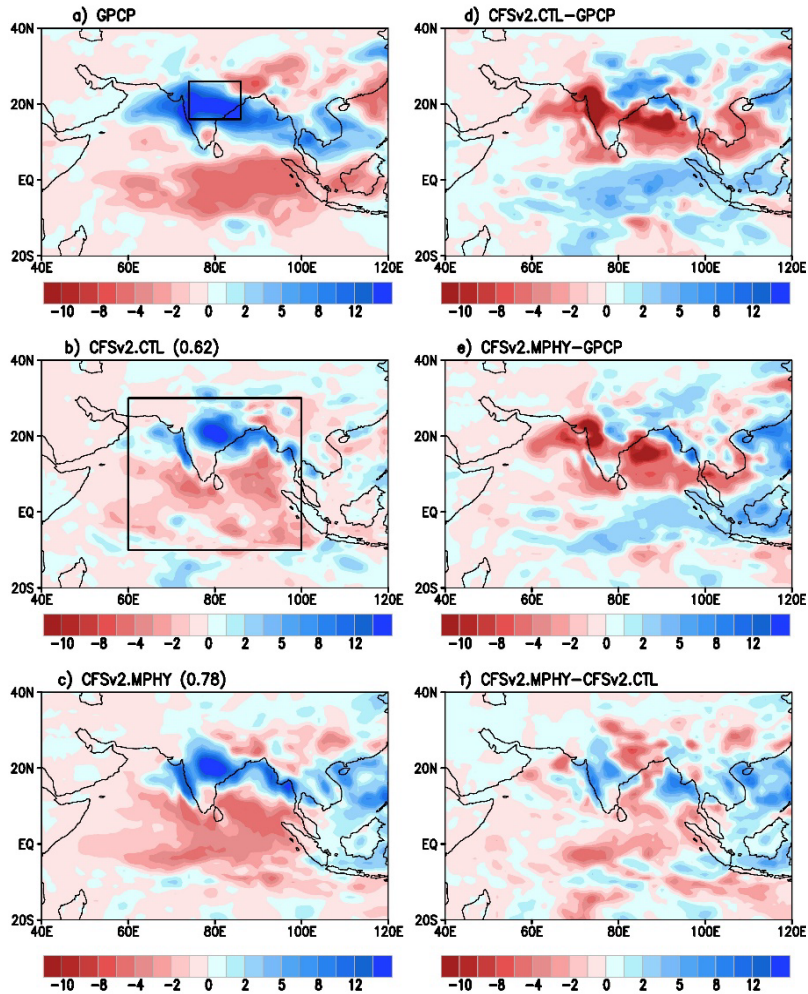

*Figure 1:* *Rainfall anomaly (mm/day) for active composite from observation (a), CFSv2.CTL (b) and CFSv2.MPHY (c). Bias in simulation for sensitivity experiments in (d) and (e). Difference in model simulation in (f). Pattern correlation of sensitivity experiments with observation over the box (shown in b) are in parenthesis.*

However, both the SE126 underestimate the spatial extent of positive rainfall anomaly over the BoB. The negative rainfall anomaly is observed over the equatorial Indian Ocean (EIO) during an active spell (Fig. 1a). The spatial extent and intensity of rainfall anomaly over EIO are better captured in CFSv2.MPHY than CFSv2.CTL (Fig. 1b, c; Table 1). The magnitude of bias of rainfall anomaly is reduced in CFSv2.MPHY over the CI region, the BoB (80 ˚E -100˚E,10˚N-20˚N), the Arabian Sea (AS: 60 ˚E -74˚E,10˚N-20˚N), and the equatorial Indian Ocean (EIO: 60˚E -110˚E,10˚S-5˚N) from CFSv2.CTL (Table 1, Fig. 1d, 1e). The spatial distribution of bias of rainfall anomaly simulation of CFSv2.CTL (Fig. 1d) shows a higher negative (positive) bias over central India (southeastern TP) than that of CFSv2.MPHY (Fig.1e). The difference between the two SE126 (Fig. 1f) demonstrates that the CFSv2.MPHY has improved rainfall simulation in most of the land regions of the Indian subcontinent. The rainfall simulation over the equatorial Indian Ocean is also improved in CFSv2.MPHY.

**Table 1. The bias of rainfall anomaly (mm/day) of CFSv2.CTL and CFSv2.MPHY for active composite with respect to observation over different regions. Similar values for break composite are in parenthesis.**

|  | CI | BoB | AS | EIO |
|---|---|---|---|---|
| **CFSv2.CTL** | -3.34 (2.67) | -4.74(1.73) | -3.14(0.58) | 2.10(-1.1) |
| **CFSv2.MPHY** | -2.75(2.52) | -3.61(-0.26) | -2.9(0.13) | 1.07(1.39) |

The break spell is characterized by rainfall below normal levels over Indian land. The below-normal rainfall activity is noticed over central India southern peninsula which also extends to the basins of the AS and the BOB (Fig. 2a). Also, during a break spell, the rainfall band shifts over the equatorial Indian Ocean (Fig. 2a). Strong negative rainfall anomaly is noticed over the BoB basin. A strong positive rainfall anomaly is noticed over the Gangetic West Bengal, extending to some parts of northeast India. The CFSv2.CTL can capture the negative rainfall anomaly over central India and near the western coast. However, it fails to simulate the observed break-spell feature over the BoB and the Gangetic West Bengal.

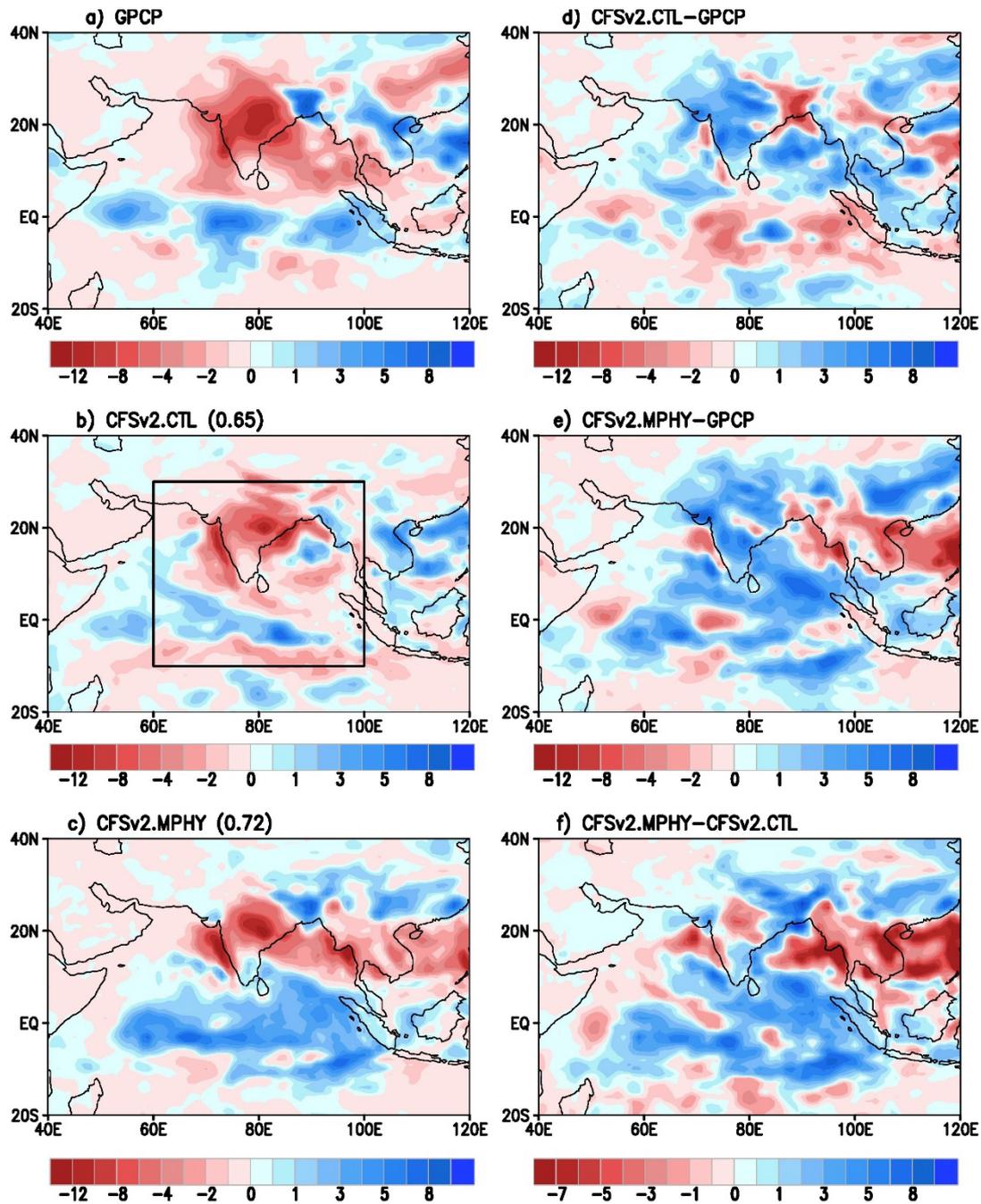

*Figure 2.* Same as Figure 1 but for break composite.

The CFSv2.MPHY (Fig. 2c) captures this feature better than CFSv2.CTL (Fig. 2b). The CFSv2.MPHY can realistically capture the distribution of the negative rainfall anomaly over the Indian subcontinent better than the control run. The simulation of positive rainfall anomaly in the central BoB by CFSv2.CTL (Fig. 2b), which sharply contrasts with observation, is now resolved in CFSv2.MPHY (Fig. 2c) Though the bias in simulating the rainfall anomaly in the break spell is still persistent in both the SE126 runs, the magnitude of bias of rainfall anomaly is reduced over the CI, the BoB, and the AS region in CFSv2.MPHY but increased for the EIO region. Additionally, the CFSv2.MPHY overestimates the positive rainfall anomaly over the EIO region, which is underestimated in the case of CFSv2.CTL (Fig. 2d, 2e, Table 1). The mean rainfall composite during active and break spells from the two SE126 runs is provided in Fig. S1 of the supplementary. The mean value of rainfall averaged over the central India region during active and break composite for observation (GPCP) and two SE126 runs are shown in Table 2. Results demonstrate that CFSv2.MPHY simulated mean rainfall closer to but somewhat overestimated than the observed value during the active period.

The OLR is considered a proxy of convection (Murakami, 1980). The deep convection (i.e., less OLR) is noticed over central India and adjacent BoB during the active spell (Fig. S2a). On the other hand, the break spell is associated with higher OLR (Fig. S2d) over the Indian subcontinent. The shifting of the convection zone is also noticed over the EIO during the break spell (Fig. S2d), as a lower value of OLR is observed over that region of the active spell (Fig. S2a). The CFSv2.CTL fails to realistically capture the spatial extent of the deep convection zone during active spell over the Indian landmass adjoining the BoB and the AS (Fig. S2b). The shift of the convection zone toward the EIO needs to be adequately captured by the CFsv2.CTL (Fig. S2e) during break spell. The sparse simulation of OLR distribution by the control run also supports its limited success in simulating the rainfall distribution during active and break spells. The progress in simulating the distribution of OLR realistically during active and break spells is noticed in CFSv2.MPHY (Fig. S2c, f) and it can realistically capture the spatial extent of a low OLR (i.e., deep convection) zone in central India and adjoining AS and BoB during active spells (Fig. S2c). However, it shows more convection over the South China Sea region, which contrasts with observation. During the break spell, the CFSv2.MPHY can capture the observed low OLR zone over the EIO and near the maritime continents (Fig. S2f). Quantitatively also1, the mean value of OLR averaged over central India demonstrated an improvement in CFSv2.MPHY from CFSv2.CTL (Table 2). CFSv2.MPHY is realistically closer to the observed value of OLR during the active spells. The mean OLR value for break composite is also better simulated in CFSv2.MPHY. The reduced $C_{CA}$ limits the conversion of cloud condensate to convective precipitation, increasing the cloudiness in the mid to upper troposphere through detrainment of moisture (J. Y. Han et al., 2016). On the other hand, the increased $C_{MA}$ increases the conversion of cloud liquid water to microphysical rain by reducing the characteristic time for the collision-coalescence process. Therefore, proper choice/combination of autoconversion coefficients can play a pivotal role in modulating the convection during the monsoon oscillation.

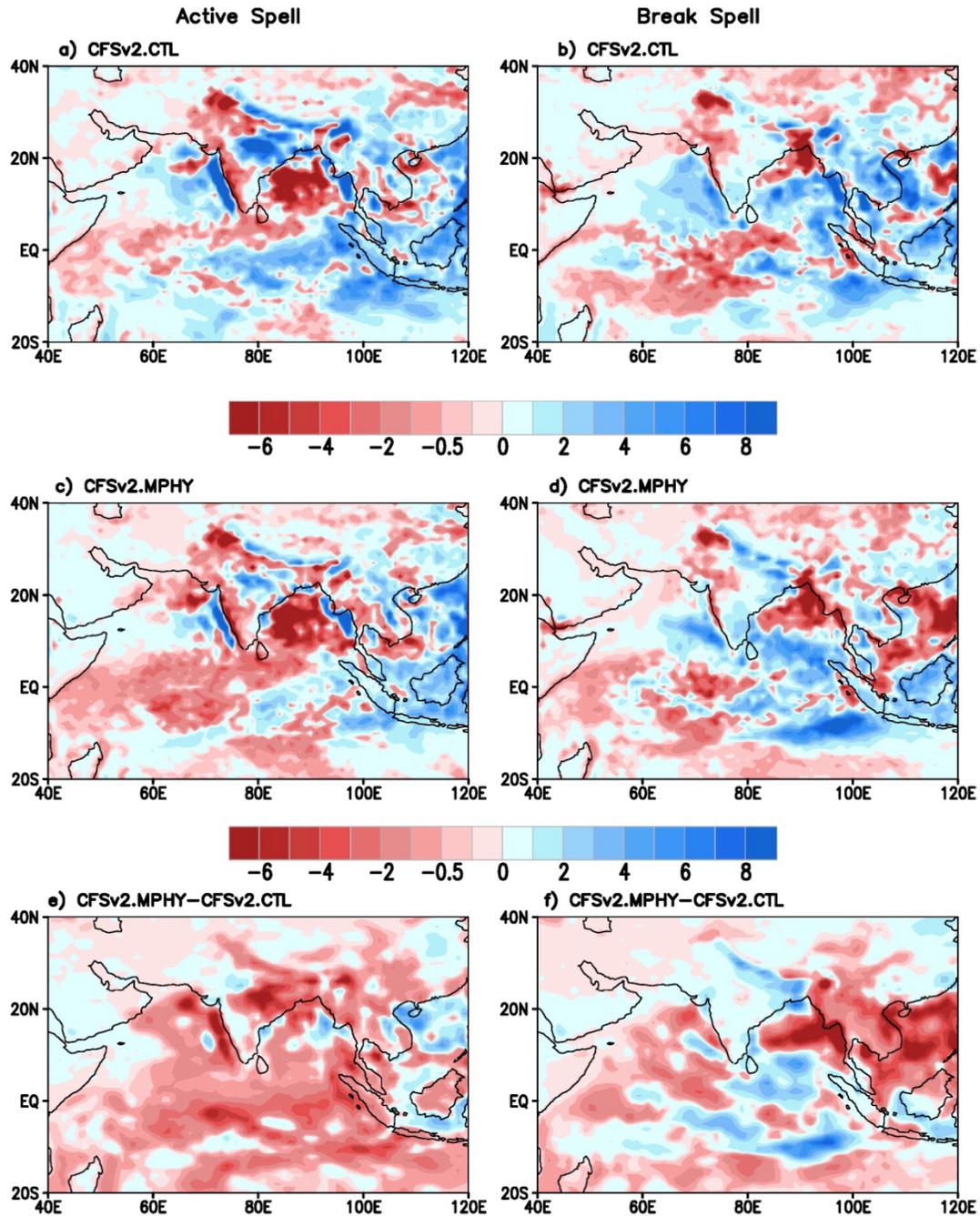

*Figure 3.* *The bias of convective rainfall for CFSv2.CTL during (a) active spell and (b) break spell. Similar bias for CFSv2.MPHY during (c) active spell and (d) break spell. The difference in convective rainfall between two sensitivity experiments, i.e., CFSv2.MPHY minus CFsv2.CTL for (e) active and (f) break spells.*

### 3.1.2. Convective Rainfall

(S. B.Saha et al., 2014) found that during active spells, precipitating clouds are more of a stratiform type (~20%) than a convective type (~5%) over the monsoon trough region. Therefore, proper representation of the convective to total rain (RCT) ratio in the global climate models is essential and requires improvement (Hazra, Chaudhari, Saha, Pokhrel, et al., 2017). In general, most of the climate models tend to simulate 95% of the rain as convective (Dai, 2006), whereas satellite observations show that 40-50% of rainfall originates from the melting of ice(Field &Heymsfield, 2015).

**Table 2. The mean value averaged over central India of different variables from Observation (or Reanalysis), CFSv2.CTL and CFSv2.MPHY for active composite. Values for break composite are in parenthesis.**

|  | Rainfall (mm/day) | OLR (Watt/m$^2$) | Convective Rain (mm/day) | RCT (%) | HCF (%) |
|---|---|---|---|---|---|
| **Observation/ Reanalysis** | 19.13(1.40) | 183.17 (235.71) | 6.53(1.11) | 48(52) | 86.28(72.31) |
| **CFSv2.CTL** | 14.19 (1.39) | 203.22 (263.53) | 8.85(1.22) | 62(83) | 63.88(19.88) |
| **CFSv2.MPHY** | 15.39 (2.88) | 185.07 (233.69) | 6.69(1.66) | 47(70) | 77.15 (50.10) |

(Pathak et al., 2019) showed that the models' RCT ratio depends on convective and microphysical parameterization. (Dutta et al., 2021), through a series of sensitivity experiments (SE), showed that a proper combination/choice of the two autoconversion coefficients could improve RCT over the Indian subcontinent on a seasonal scale. However, the status of RCT during active and break spells was not evaluated. We have compared the convective rainfall simulated by the SE126 runs with that of reanalysis (ERA5). We have considered the reanalysis for comparison as satellite-derived bifurcation of convective and stratiform components of rainfall is classified differently than a model (Ganai et al., 2019). The bias of convective rainfall of both SE126 runs is shown for active and break spells (Fig. 3). The CFSv2.CTL strongly overestimates the convective rainfall over central India and Himalayan foothills during the active spell (Fig. 3a), which is considerably reduced in CFSv2.MPHY (Fig. 3c). The wet bias of convective rainfall along the western coast of India over the AS is also reduced in CFSv2.MPHY (Fig. 3c) from CFSv2.CTl (Fig. 3a). However, both the SE126 runs underestimate the convective rainfall over the BoB. Interesting results are seen over the equatorial Indian Ocean (EIO) during the active spells. At the same time, the wet bias over the eastern EIO near the maritime continents is reduced in CFSv2.MPHY (Fig. 3c), the dry bias over the western EIO is more prominent in CFSv2.MPHY than CFSv2.CTL (Fig. 3a). Prominent differences, such as active spells, are not seen between the SE126 runs during break spells. The spatial extent of wet bias over the AS is reduced in CFSv2.MPHY (Fig. 3d), but the dry bias over the BoB covers a more spatial area in CFSv2.MPHY than that of CFSv2.CTL (Fig. 3b). A similar bias distribution is noticed for both the SE126 runs over the EIO, with less magnitude in CFSv2.MPHY than CFSv2.CTL is noticed for the break spells. The difference in the convective component of rainfall between the two SE126 runs is also demonstrated to better portray the improvement of the modified run. Results show that convective rainfall is reduced in CFSv2.MPHY during active and break spell over most parts of the basin from CFSv2.CTL (Fig. 3e, f). However, during the break spell, the convective rainfall is simulated slightly more over the Indian landmass by the CFSv2.MPHY (Fig. 3f). Quantitatively, the mean value of convective rainfall is more (less) over central India during active composite (Table 2) in CFSv2.CTL (CFSv2.MPHY). The situation is reversed during the break spell (Table 2). The mean value of convective rain for CFSv2.MPHY is considerably closer to the reanalysis for an active spell, whereas CFSv2.CTL is slightly closer during the break spell

The RCT from reanalysis (ERA5) for active and break spells is provided in the supplementary figure (Fig. S3a, d). Results show that the RCT simulated by CFSv2.MPHY (Fig. S3c) during active spell over the Indian subcontinent (land region) is also better than CFSv2.CTL (Fig. S3b). Over the oceanic part, both the SE126 runs overestimate the RCT, which is eventually reduced in CFSv2.MPHY. For break composite, both the SE126 runs (Fig. S3e, f) show high RCT (>90%) over most of the oceanic parts of the Indian subcontinent, which is in sharp contrast with the reanalysis (Fig. S3d). The RCT values over the western part of the EIO are realistically improved for CFSv2.MPHY (Fig. S3f). Over the landmass, both the SE126 runs overestimate the RCT. However, CFSv2.MPHY can realistically capture the low RCT over central India for the break spell. Quantitatively, from Table 2, we see that RCT values are simulated by CFSv2.MPHY is closer to the reanalysis for both active and break spells over the central India region than CFSv2.CTL. Since the high (low) value of $C_{CA}$ and $C_{MA}$ increases (decreases) the convective and large-scale rainfall, the results suggest proper modulation of autoconversion coefficients can improve the RCT realistically in the modified run from control.

### 3.1.3. Clouds and Circulation

(Chaudhari et al., 2016) have shown that clouds impact rainfall variability through the modulation of heating (Baker, 1997; Y. Hong et al., 2016) and induced circulation (Kumar et al., 2014). Reanalysis shows that the extensive amount of HCF (>90%) covers most of the Indian landmass along with adjoining As and BoB during active spells (Fig. 4a). The HCF distribution of HCF in the active spell is also consistent with the convection zones (Fig. S2a) as much presence of cloud in mid to high tropospheric levels trap the OLR more. The CFSv2.CTL critically underestimates the HCF over the Indian subcontinent during the active spell (Fig. 4b, Fig. S4a). The distribution of HCF during active spells is improved in CFSv2.MPHY (Fig. 4c) and the bias of HCF simulation during active composite is considerably reduced in CFSv2.MPHY (Fig. S4c). The BoB is considered a convective heating region where many low-pressure systems (LPS) form during monsoon and traverse westward as a major rain-bearing mechanism(Goswami, 1987; Goswami et al., 2003).

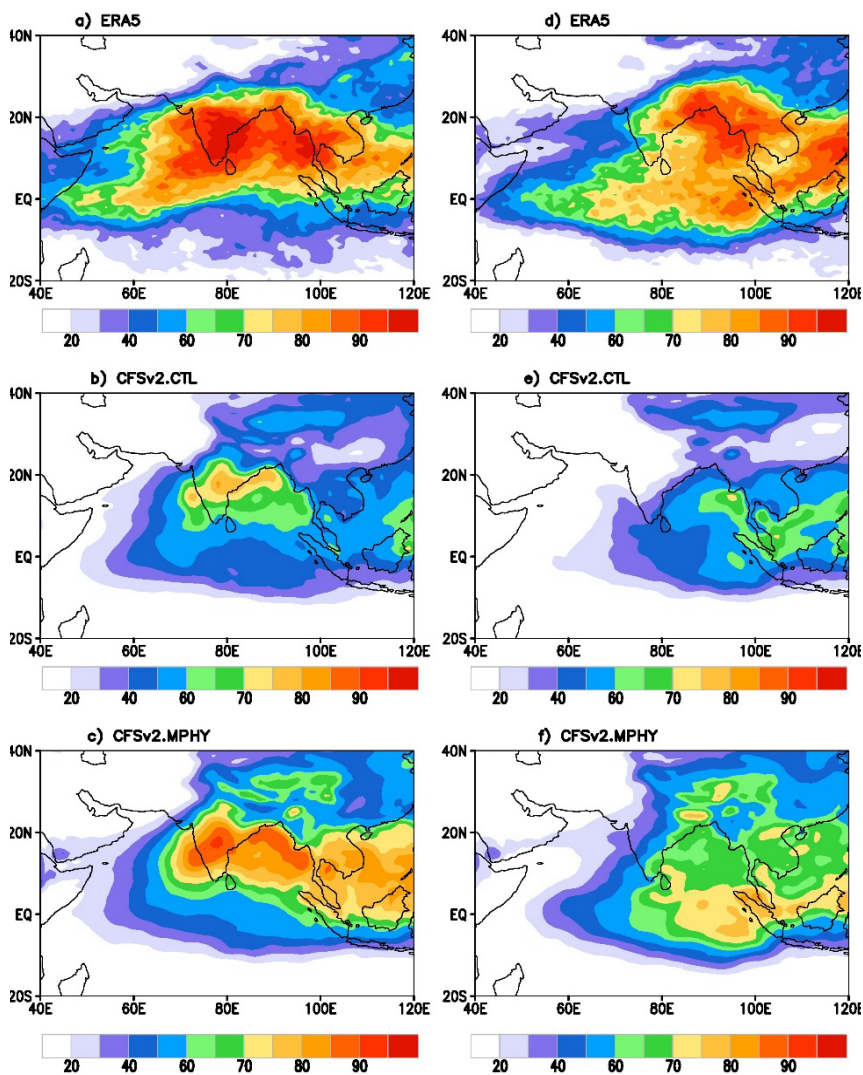

*Figure 4.* *Mean value of high cloud fraction (%) for active (a-c) and break (d-f) composite from reanalysis (a, d), CFSv2.CTL (b, e), and CFSv2.MPHY (c, f).*

Therefore, abundant amounts of high clouds are seen over the BoB for both active and break spells (Fig. 4a, d), being more in an active spell. The high negative bias is shown by CFSv2.CTL (Fig. S4a) over this BoB region in the active spell, which is well reduced in CFSv2.MPHY (Fig. S4c). The HCF amount is well reduced over the Indian landmass from active spell to break spell as shown by the reanalysis (Fig. 4d). The reduction of HCF is also captured by both the SE126 runs. However, the bias in simulation is also improved in CFSv2.MPHY (Fig. S4d) from CFSv2.CTL (Fig. S4b). Moreover, the CFSv2.MPHY can realistically capture the increase of HCF near the maritime continents over the EIO. The quantitative calculation of HCF over central India during active and break spells from the reanalysis and two SE126 runs confirms the CFSv2.MPHY performs better in simulating the HCF than that of CFSv2.CTl (Table 2). The improvement in HCF simulation by the CFSv2.MPHY has also been reflected in simulating the OLR distribution (Fig. S2c, f).

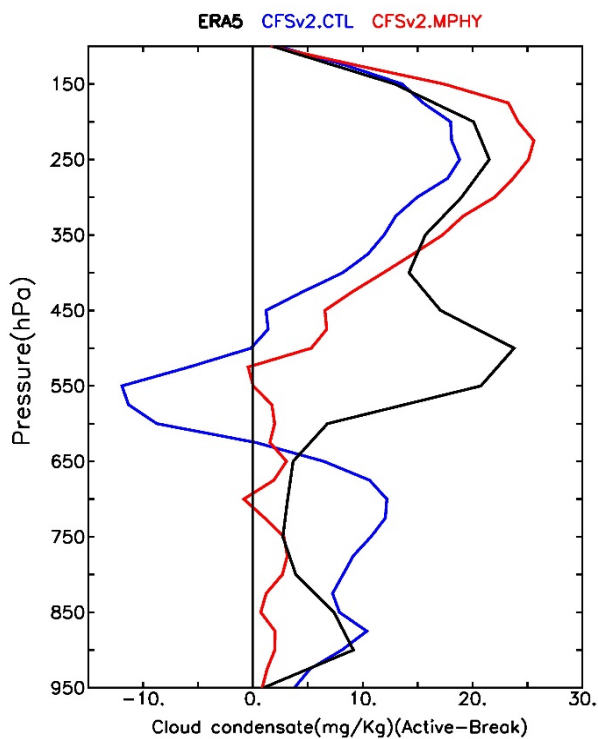

*Figure 5.* *The difference in the simulation of the vertical profile of cloud condensate averaged over the All-India region (70˚E - 90˚E, 10˚N-30˚N) between active and break composite from reanalysis (ERA5), CFSv2.CTL and CFSv2.MPHY.*

(Dutta et al., 2020) showed that the upper-level cloud condensate over the ISM region in an active spell is much higher than in a break spell. To understand the improvement in cloud profile in CFSv2.MPHY, during the intraseasonal fluctuation of ISMR, we compared the cloud condensate simulation of SE126 runs (Fig.5) with a reanalysis (ERA5). The sum of cloud ice water and cloud liquid water is considered as cloud condensate for reanalysis. The difference of cloud condensate between active and break composite shown for reanalysis and both the SE126 runs. In the lower tropospheric region (850 hPa to 650 hPa), the CFSv2.MPHY is closer to the reanalysis. The results show that in the mid-troposphere, the CFSv2.CTL shows more cloud condensate during the break spell, which is realistically reversed for CFSv2.MPHY, but the difference is still underestimated. The difference in cloud condensate in the upper troposphere is more (less) in CFSv2.MPHY (CFSv2.CTL) than reanalysis. A reduced $C_{CA}$ leads to weaker precipitation, leaving more cloud water in the convective precipitation and more cloud water detrainment into the stratiform (grid-scale) clouds. By contrast, the increased $C_{MA}$ shortens the characteristic time of the warm rain autoconversion process, strengthening the stratiform precipitation. Therefore, the proper choice of autoconversion parametrization can lead to better simulation of the mean vertical structure of the atmosphere and dynamic field (Dutta et al., 2021; Ganai et al., 2019; J. Y. Han et al., 2016; Hazra et al., 2023).

The autoconversion strength also affects the regional (averaged over 70° E–90° E) Hadley circulation (expressed in terms of pressure vertical velocity (omega). During the active condition, reanalysis shows a strong ascending branch (Fig. 6a) over Indian latitudes (i.e., 10°N-30°N), which is weakened during the break spell (Fig. 6d). On the other hand, a stronger ascending branch over the south of the equator is noticed for break spell (Fig. 6d) than active spell (Fig. 6a). CFsSv2.CTL captures a strong ascending branch around 20°N during the active condition (Fig. 6b) and over the south of the equator during the break spell (Fig. 6e). CFSv2.MPHY shows a strong ascending branch over the north of the equator to 30°N during the active spell (Fig. 6c), which is weakened during the break spell (Fig. 6f). The spatial extent of the ascending/descending branch of CFSv2.MPHY is in better agreement with that of reanalysis. This improvement also reflects the cloud condensate vertical profile in Fig. 5 and the high cloud fraction over central India (Table 2). The values of HCF are critically underestimated in CFSv2.CTL. The HCF simulated by CFSv2.MPHY is realistically closer to the reanalysis for active and break spells (Table 2). (Dutta et al., 2021) showed that improved bifurcation of convective and stratiform rain leads to an improved vertical profile of diabetic heating. Therefore, the "autoconversion" can modify dynamics (Hadley circulation, Fig. 6) and temperature profile (Figure not shown) through a change in heating, which in turn improves the ISM intraseasonal characteristics.

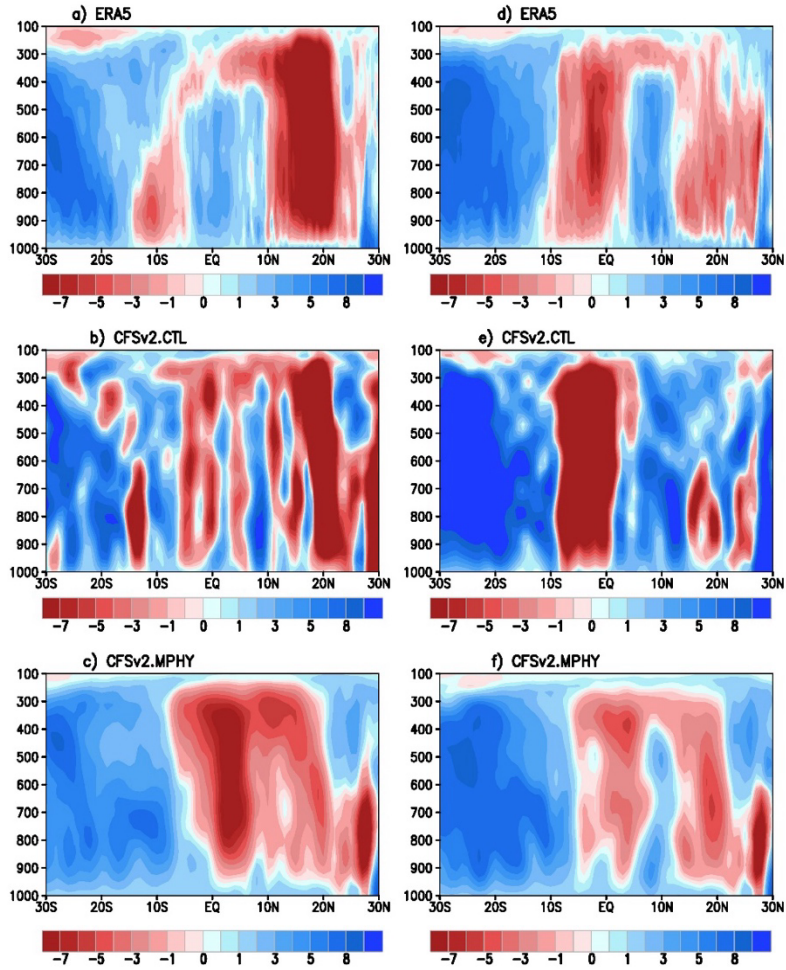

*Figure 6: Hadley circulation for active (a-c) and break composite (d-f) from reanalysis and sensitivity experiments. Pressure vertical velocity (Omega) is considered. Omega values (Pa/S) are multiplied by 100 for a better view. Positive omega values denote the descending branch of the Hadley circulation, and negative values denote the ascending branch.*

## 3.2. Subseasonal Variability

To assess the sensitivity experiments in simulating the subseasonal variabilities, we have computed the variance of rainfall (Fig. 7) for the following modes: a) Synoptic (3-7 days), b) Quasi-Bi-weekly Mode (10-20 days) and c) Low-Frequency MISO (30-60 days mode). A high variance of rainfall is observed in synoptic mode (Fig. 7a) over the Indian landmass and the EIO, which is, in general, underestimated by both the SE126 runs with some exceptions, for example, near the maritime continents. However, the variance simulated by both the SE126 runs is realistic as revealed by their PCC (Fig.

7d, g). Quantitatively, the synoptic variance (Table 3) over the Extended Indian Monsoon Region (EIMR: 70˚E-110˚E,10˚N-30˚N) and All-India (AI: 70˚E-90˚E,10˚N-30˚N) region shows that CFSv2.MPHY is closer to the observed value than that of CFSv2.CTL. However, both the SE126 runs underestimate the observed variance.

The peculiarity of rainfall simulation by recent generation models over land and ocean for ISMR has been well documented in earlier studies (Choudhury et al., 2022; Gusain et al., 2020; Jain et al., 2019). The models tend to overestimate the rainfall over the ocean, whereas they underestimate the land (Dutta et al., 2022). Here, we also tried to see how the model behaves to capture the observed rainfall variance for land and ocean separately over the whole basin (40˚E-120˚E,15˚S-40˚N). Observation (Table 3) shows that synoptic variance over land is slightly higher (~ 1 mm$^2$/day$^2$) than over the ocean—the magnitude of synoptic variance simulated by CFSv2.MPHY is closer to the observed value for each case than CFSv2.CTL. However, both the SE126 runs contrastingly show higher variances over oceans than land. Interestingly, the absolute difference in simulated variance between land and ocean parts is reduced in CFSv2.MPHY (~ 7.2 mm$^2$/day$^2$) from CFSv2.CTL (~ 9 mm$^2$/day$^2$).

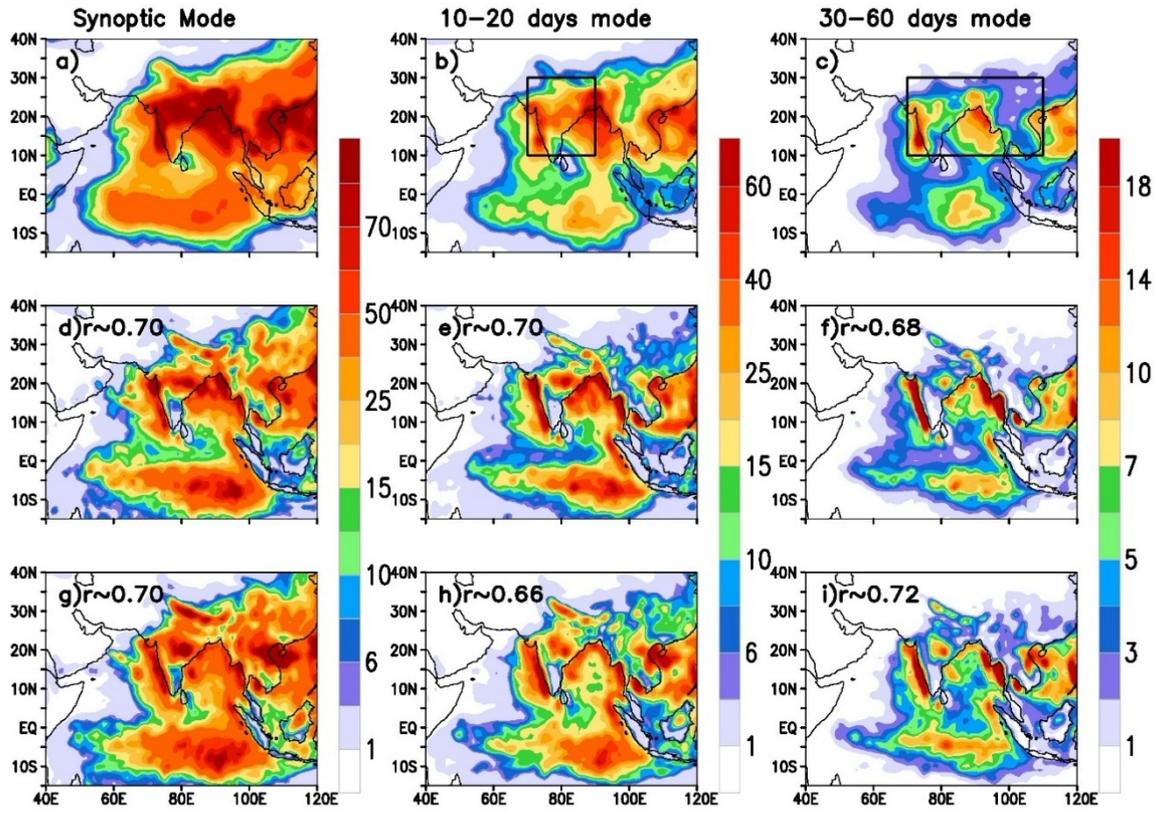

*Figure 7. Variance of rainfall (mm²/day²) for synoptic (3-7 days), quasi-biweekly (10-20 days), and low frequency (30-60 days) subseasonal mode. The pattern correlation coefficient of the sensitivity experiments with the observed variance over the whole basin is written in the respective panels.*

The rainfall variance for quasi biweekly mode or 10-20 days mode (Hazra et al., 2020) from observation shows high variance over the Western Ghats, Central India, and adjacent Bay of Bengal (fig. 7b). The magnitude of variance is less compared to synoptic mode. Models can considerably capture the mean pattern with slightly different PCC (Fig. 7e, h). A similar quantitative analysis of variance for this mode yields that both the SE underestimates the mean-variance over EIMR and AI (Table 3). Interesting results are seen when the land and ocean are separately analyzed. Unlike the synoptic mode, the observed variance is higher over the ocean (~ 2 mm²/day²) than on land. CFSv2.CTL and CFSv2.MPHY also shows the observed variance higher over ocean than land. However, the magnitude of the difference is higher than the reality, which is improved in CFSv2.MPHY (Table 3).

**Table 3. Averaged variance of rainfall (mm$^2$/day$^2$) for different modes of subseasonal oscillations.**

|  | Basin (Land Only) | Basin (Ocean Only) | Difference (Land-Ocean) | EIMR | AI |
|---|---|---|---|---|---|
| **Synoptic Mode (3-7 days)** | | | | | |
| **GPCP** | 27.32 | 26.21 | 1.11 | 63.58 | 59.7 |
| **CFSv2.CTL** | 12.26 | 21.16 | -8.90 | 33.08 | 33.32 |
| **CFSv2.MPHY** | 16.39 | 23.66 | -7.27 | 40.03 | 37.59 |
| **Quasi-Bi-weekly Mode (10-20 days)** | | | | | |
| **GPCP** | 11.01 | 12.88 | -1.87 | 28.35 | 27.54 |
| **CFSv2.CTL** | 6.231 | 15.4 | -9.17 | 22.34 | 24.19 |
| **CFSv2.MPHY** | 7.313 | 15.48 | -8.17 | 21.53 | 22.24 |
| **Low Frequency (30-60 days) Mode** | | | | | |
| **GPCP** | 2.11 | 3.31 | -1.20 | 5.88 | 6.30 |
| **CFSv2.CTL** | 1.31 | 3.97 | -2.66 | 5.67 | 5.59 |
| **CFSv2.MPHY** | 1.56 | 4.22 | -2.66 | 5.65 | 5.74 |

Observation shows that the variance of the low-frequency mode of MISO (30-60 days mode) is less than the other two modes (Fig. 7c), which the SE also captures. Both the SE126 runs overestimate the observed variance over the Western Ghats and the east coast of the Bay of Bengal (Fig. 7f, i). However, the mean value of the simulated variance over EIMR and AI aligns with the observation (Table 3). In this mode, the observed variance over the ocean is more than over land, as shown by both SE126 runs. The results from the subseasonal variance analysis show that proper modification of the autoconversion process leads to a better representation of subseasonal variability. In particular, the synoptic mode of variability shows considerable enhancement from CFSv2.CTL to CFSv2.MPHY. This is intriguing since the synoptic mode of variability influences the active/break spells (Goswami et al., 2003) and strongly correlates with mean ISMR (S. K. Saha et al., 2019). Moreover, once considered chaotic, the synoptic components of rainfall were recently found to be predictable (S. K. Saha et al., 2019, 2020).

(Suhas et al., 2013) presented a 'real-time monitoring index' for better identification of the amplitude and phase of MISO over the ISM region. They also demonstrated that these MISO indices are useful for quantifying the skill of extended-range forecasts of MISOs. Here, we demonstrate the normalized MISO index from observation and two sensitivity experiments. Normalized Power vs. Frequency (day$^{-1}$) plots for both the MISO index are shown for observation and two sensitivity experiments (Fig. 8). MISO-1 and MISO-2 for observation show peaks at ~ 0.02 day$^{-1}$ (i.e., 50 days). CFSv2.CTL simulates both peaks at ~ 0.01 day$^{-1}$ (i.e., 100 days), which is remarkably improved in CFSv2.MPHY. CFSv2.MPHY shows the peak at around 60 days and the power spectrum of MISO-1 and MISO-2 in CFSv2.MPHY is more realistic.

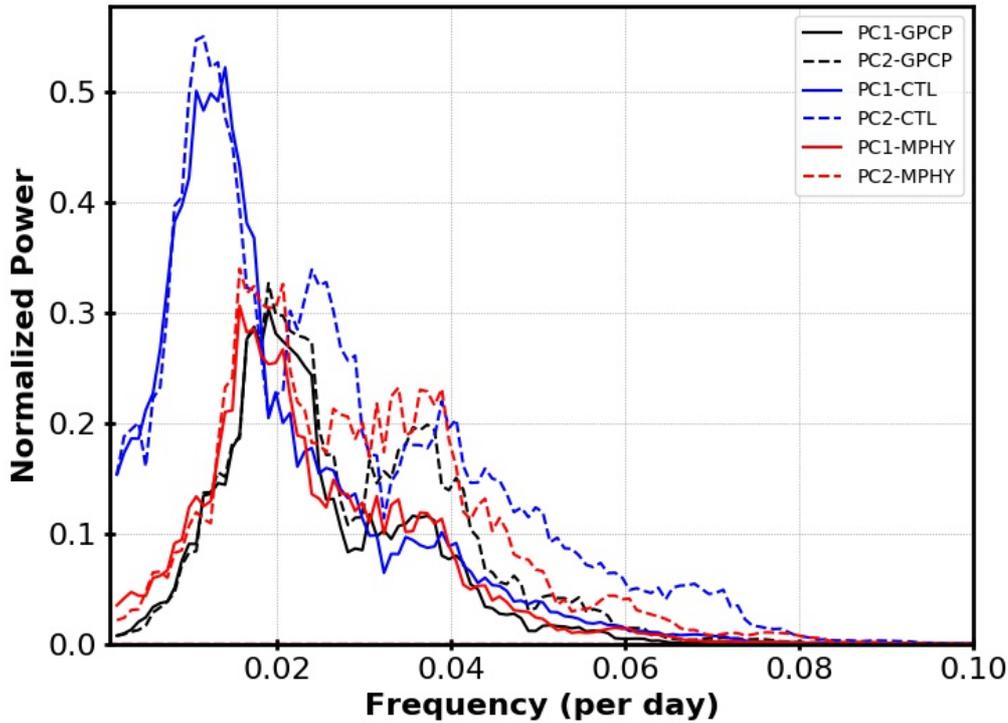

**Figure 8.** Normalized power of MISO Indices from Observation and Sensitivity experiments.

**4. Results from high-resolution Sensitivity experiments**

The results for active and break spell simulation of rainfall from SE382 runs are shown in Figure 9. The SE382 runs can better capture the observed spatial pattern (Fig. 1a) of the rainfall over the Indian subcontinent (both land and ocean part) during active spell qualitatively (Fig.9 a, b) and quantitatively (Fig. 10a). However, over the EIO the observed negative rainfall anomaly is underestimated in the SE382 than CFSv2.MPHY (Fig. 10a). The difference in simulating the active spell between the SE382 (Fig.9c) shows that CFSv2.LD-HR simulates more(less) rainfall over the southern(northern) part of the Indian landmass than CFSv2.SQ-HR. However, a quantitative average over EIMR (land only) yields that CFSv2.SQ-HR is the closest to the observed value for an active spell among all the sensitivity experiments (Fig. 10a), followed by CFSv2.LD-HR. Interesting results are seen for the break spell (Fig. 9d-f).

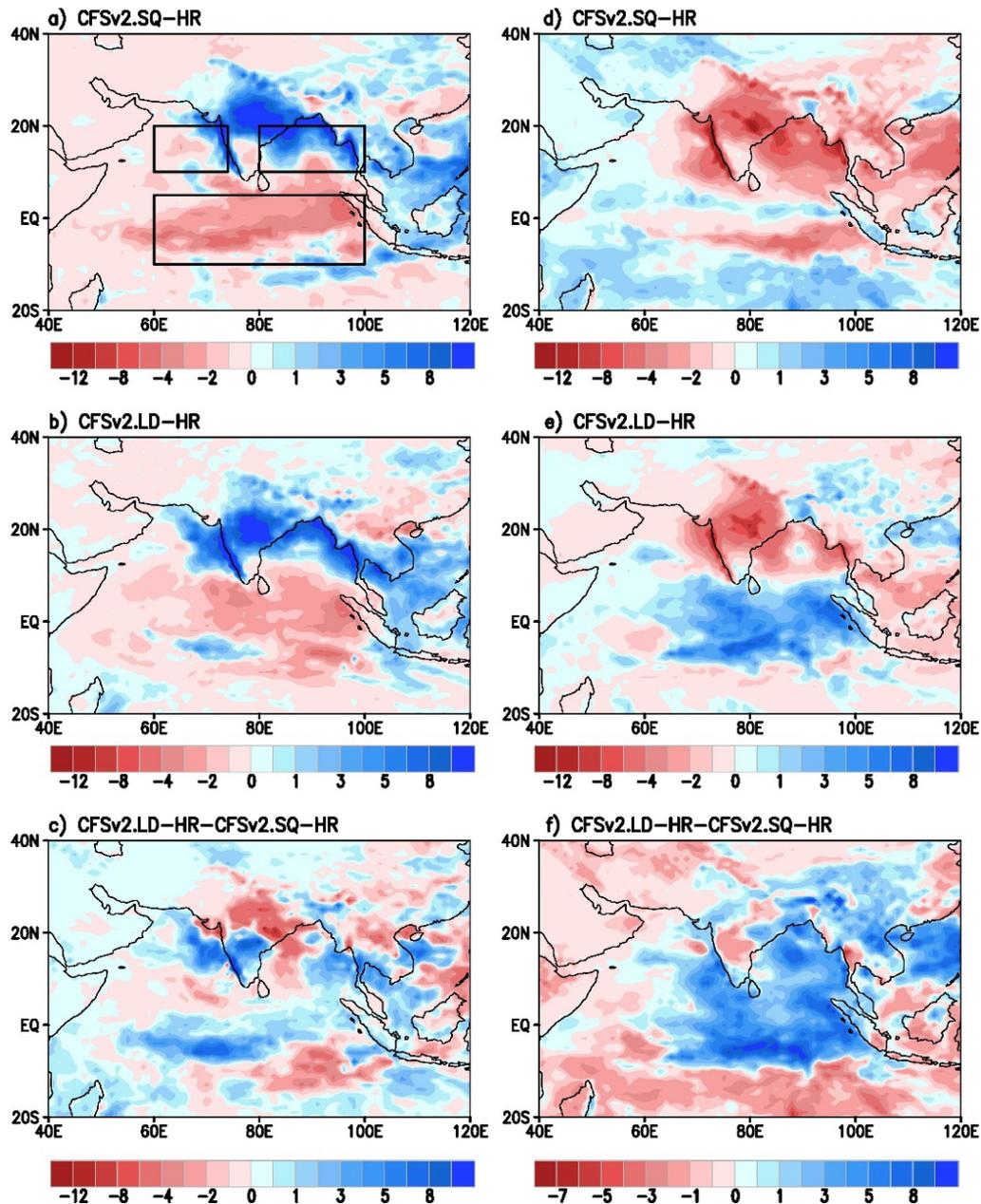

***Figure 9:*** *Rainfall anomaly (mm/day) for active composite from (a) CFSv2.SQ-HR and (b) CFSv2.LD-HR and their difference in the simulation of the active composite are in (c). Rainfall anomaly for break composite is in (d) CFSv2.SQ-HR and (e) CFSv2.LD-HR. Differences in the simulation of the break composite are in (f). AS, BoB and EIO regions are shown in a.*

The CFSv2.LD-HR is realistically closer than its counterpart to the observed (Fig. 2a) break spell. The rainfall over the oceanic part of the Indian subcontinent, e.g., AS, BoB, EIO, is better simulated in CFSv2.LD-HR (Fig. 9e). The CFSv2.SQ-HR unrealistically simulates less rainfall over these oceanic parts than CFSv2.LD-HR (Fig. 10b, 9f).

Quantitative comparison among all the sensitivity experiments (Fig. 10b) shows that over the EIO, only the CFSv2.MPHY and CFSv2.LD-HR can capture the proper sign of rainfall anomaly (i.e., positive). The overestimation of negative rainfall anomaly over the EIMR (land only), AS, and BOB is also noticed in CFSv2.SQ-HR. CFSv2.LD-HR and CFSv2.MPHY yields better results in this regard (Fig. 10b).

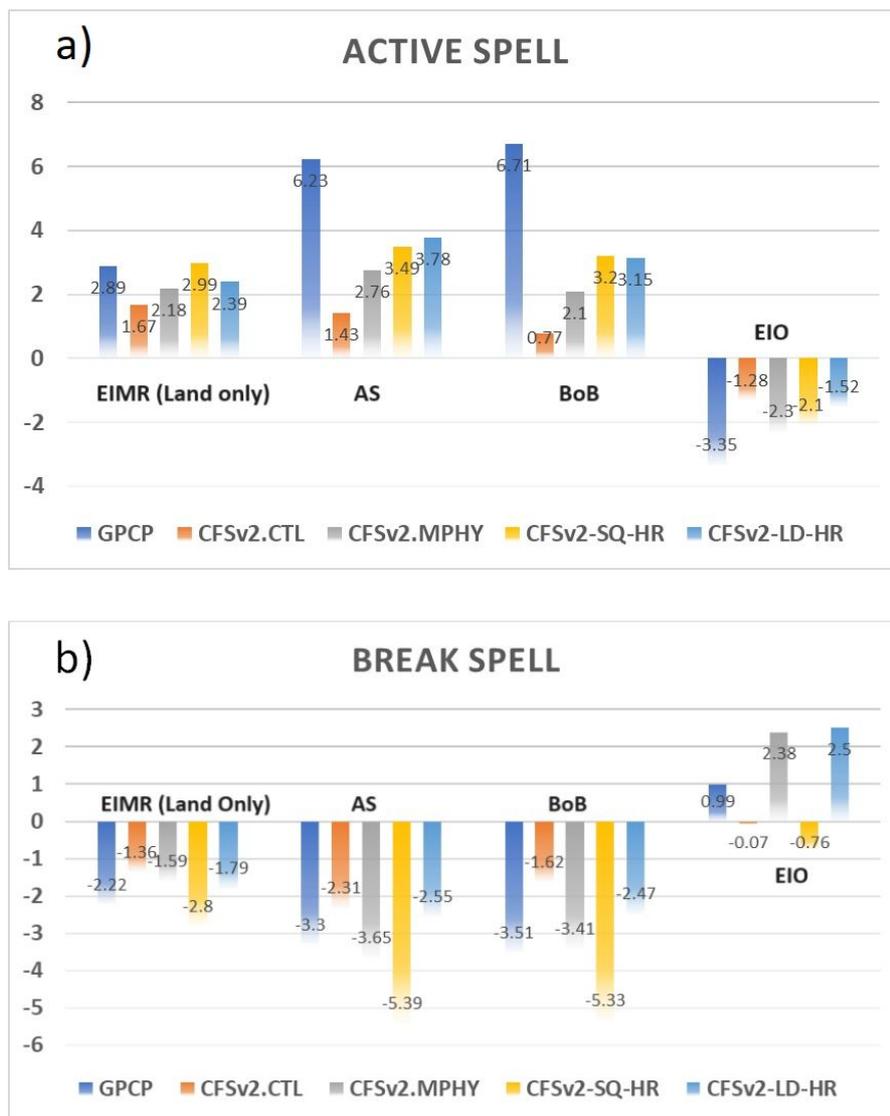

*Figure 10. Quantitative estimate of rainfall anomaly over selective regions for a) active spell and b) break composite from observation and all sensitivity experiments considered in the study.*

Normalized Power vs. Frequency (day$^{-1}$) plots for both the MISO index are shown for observation, and two sensitivity experiments with high-resolution CFSv2 (CFSv2.SQ-HR and CFSv2.LD-HR) are also shown in supplementary Figure S5. CFSv2.LD-HR simulates both peaks better than CFSv2.SQ-HR. Moreover, the power spectrum of MISO-1 and MISO-2 in CFSv2.LD-HR agrees with the observation better than CFSv2.SQ-HR.

Several studies have demonstrated the dominant northward propagation over the ISM region of the low-frequency MISO. To examine the fidelity in simulating this aspect, finite domain space-time spectra (Wheeler &Kiladis, 1999) of 20–100 days filtered rainfall anomalies (averaged over 60°E - 110°E) are computed from observations (GPCP and TRMM, Figures 11a and 11b) and the four sensitivity experiments (e.g., CFSv2-CTL, CFSv2.MPHY, and CFSv2-SQ.HR, and CFSv2.LD-HR; Figures 11c-f). The maximum intensity of MISO is observed at wave number one and period of 40 days (Figs. 11a,b), but the CFSv2-CTL and CFSv2.SQ-HR simulates it with higher (> 40 days) and lower (< 40 days) periods (Figure 11c, e). The CFSv2.MPHY, and CFSv2.LD-HR (Figure 11d, f) experiments simulate the peak intensity at meridional wave number one and period 40 days, which aligns with both observations (GPCP and TRMM). These findings pinpoint that proper autoconversion parameterization is required to improve the simulation of the observed space-time structure of the MISOs. Our future study will explore the physical reasoning behind these improvements in the SE382.

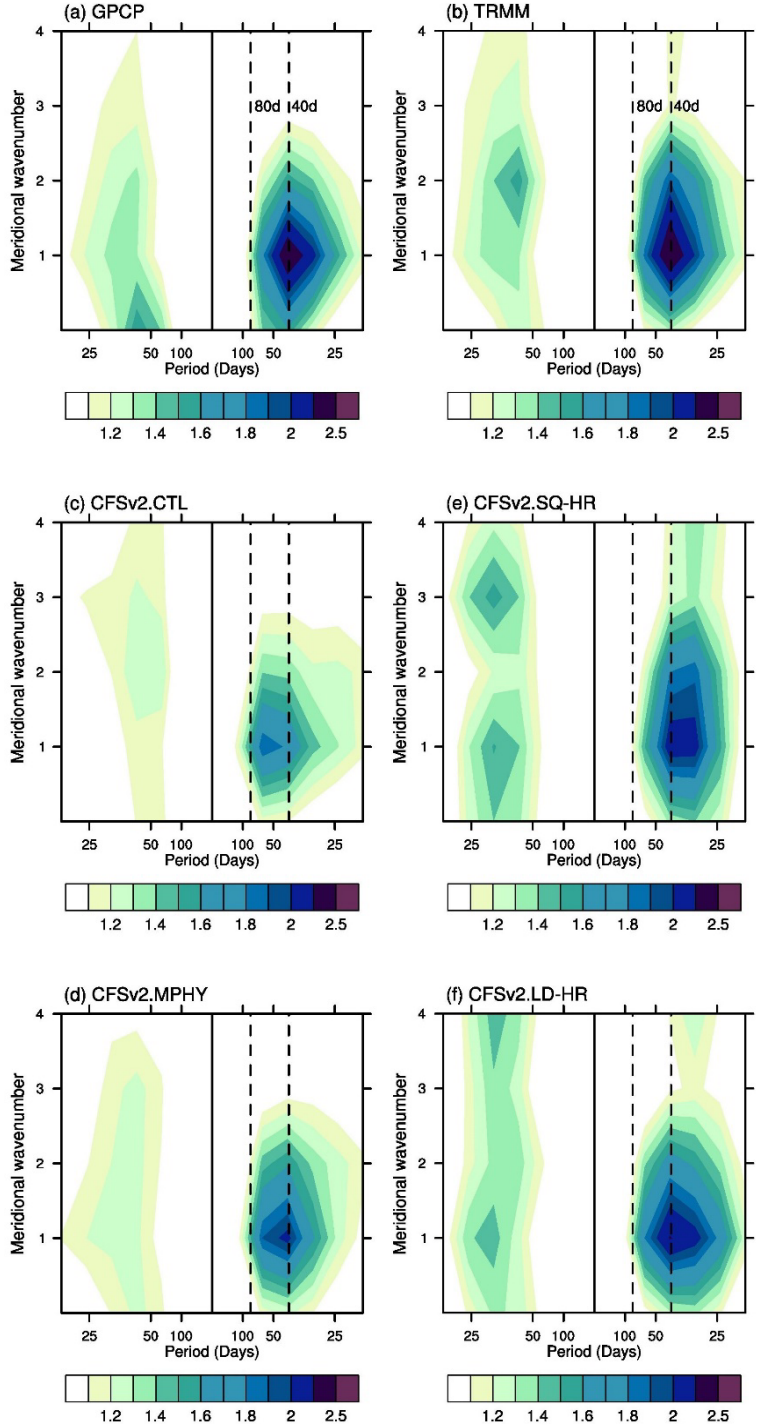

*Figure 11. The finite domain space-time spectra of 20–100 filtered rainfall averaged over 60˚E–110˚E computed from observations (a. GPCP; b. TRMM) and the four experiments (c. CFSv2.CTL; d. CFSv2.MPHY; e. CFSv2.SQ-HR; f. CFSv2.LD-HR).*

## 5. Summary

The study examines the representation of rainfall patterns during active and break spells of the Indian Summer Monsoon (ISM) using observations and two sets of sensitivity experiments (SE126 and SE382) with different resolutions of the CFSv2 model. In SE126 runs, the microphysical autoconversion was based on Sundqvist et al. (1989), and sensitivity experiments were performed in CFSv2 with a horizontal spectral resolution of ~ 100 km (T126). In SE382 runs, the sensitivity experiments were performed in CFSv2 with a horizontal spectral resolution of ~ 38 km (T382) with a control run using the Sundqvist scheme and a modified one using the Liu-Daum autoconversion scheme.

The main findings from the study are summarized below.

a) During active spells, observations show anomalously heavy rainfall over most of India, the Bay of Bengal (BoB), the Western Ghats, and the Arabian Sea (AS). The equatorial Indian Ocean (EIO) experiences less rainfall. Both SEs capture these patterns to varying degrees, but CFSv2.MPHY improves the spatial extent and intensity of rainfall anomalies over central India and the EIO compared to CFSv2.CTL. The break spell is characterized by below-normal rainfall over India and a shift of the rainfall band to the EIO. This pattern is better simulated by CFSv2.MPHY.

b) Both SE126 runs underestimate rainfall over central India, BoB, and AS, with CFSv2.CTL shows greater underestimation. CFSv2.MPHY shows improved rainfall simulation across most land regions and the EIO.

c) Proper representation of the ratio of convective to total rain (RCT) is crucial. During active spells, CFSv2.MPHY shows a lower RCT, aligning with observed stratiform precipitation dominance.

d) HCF is observed higher during active spells over the Indian subcontinent. CFSv2.MPHY improves the simulation of HCF and vertical cloud condensate profiles compared to CFSv2.CTL. The Hadley circulation patterns in CFSv2.MPHY also aligns better with reanalysis data.

e) For Synoptic Mode (3-7 days), high variance over Indian landmass and EIO are underestimated by both SE126 runs, but CFSv2.MPHY is closer to observed values. In Quasi-Biweekly Mode (10-20 days), there is high variance over the Western Ghats, central India, and BoB. Both SE126 runs capture the pattern but underestimate the variance with CFSv2.MPHY showing closer agreement to observations. The observed variance in the low frequency (30-60 days) mode is less than the other two modes. Both SE126 runs overestimate variance over the Western Ghats and BoB but align well with the observed values for EIMR and AI regions.

f) CFSv2.MPHY shows an improved representation of the MISO index, with peaks in normalized power closer to observed values, indicating better simulation of the MISO periodicity.

g) The preliminary analysis with the SE382 runs suggests that Liu-Daum-based autoconversion in a high-resolution set-up of CFSv2 has the potential for better representation of active-break spells over the Indian subcontinent.

h) Modification of autoconversion coefficients in the SE126 category and using the Liu-Daum autoconversion scheme instead of the Sundqvist scheme in the SE382 category resulted in improved representation of the northward propagation feature of MISO.

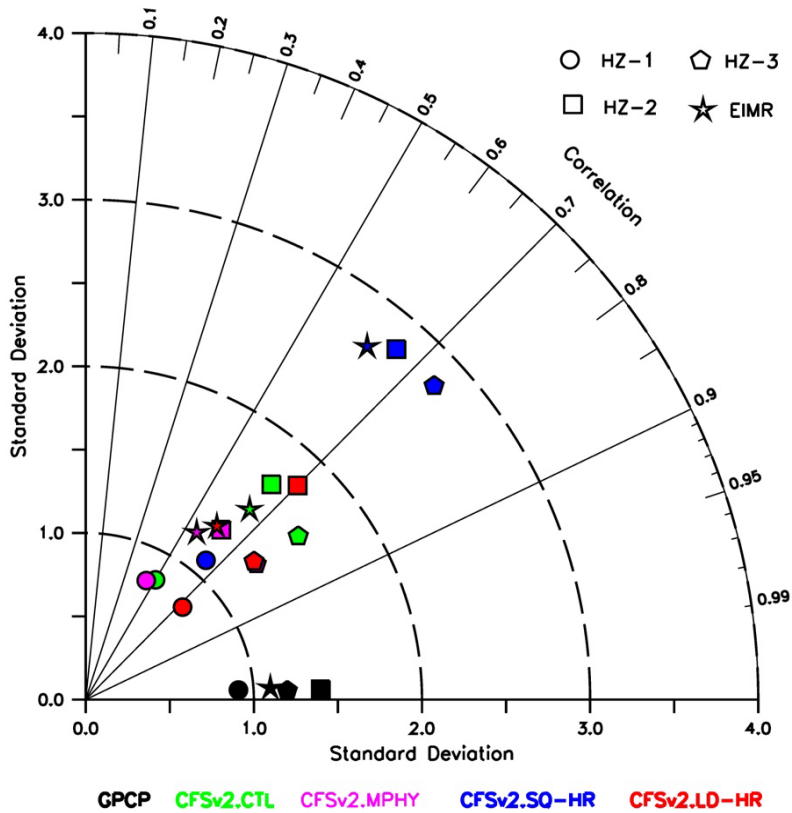

*Figure 12. Taylor plot showing interannual standard deviation (mm/day) for rainfall (JJAS) from observation (GPCP) and the four experiments CFSv2.CTL, CFSv2.MPHY, CFSv2.SQ-HR, and CFSv2.LD-HR and the PCC of the sensitivity experiments with observation over different homogeneous zones (HZ) and EIMR.*

For a quantitative comparison of simulation fidelity among sensitivity experiments, simulated standard deviation (interannual) of ISM rainfall averaged over different homogeneous zones (HZ), e.g., HZ-1 (68°E - 89°E, 25°N - 33°N); HZ-2 (68°E - 89°E, 17.5°N - 25°N); HZ-3(68°E - 89°E, 8°N – 17.5°N) and EIMR are compared with observation along with respective PCC over those regions (Fig. 12). Results show that CFSv2.SQ-HR largely overestimates the standard deviation (SD) over HZ-2, HZ-3, and EIMR. In the case of SE126 runs CFSv2.MPHY is better than CFSv2.CTL for most of the cases. Among all the sensitivity experiments, CFSv2.LD-HR is better regarding the simulation of interannual SD and PCC. Therefore, this study highlights that proper autoconversion parameterization in the coupled climate model can enhance the representation of active/break spells and sub-seasonal variability of ISM.

**Acknowledgements:**

UD thanks the funding support from the National Science and Technology Council of Taiwan under grant NSTC 112-2811-M-002-131. MB, AH, and SAR thank MoES, the Government of India, and the Director of IITM. AH also acknowledges the National Science and Technology Council (NSTC), Taiwan, for funding support as a visiting researcher at National Taiwan University, Taiwan. We also thank the freely available software viz. The Grid Analysis and Display System (GrADS), NCAR Command Language (NCL), Ferret-NOAA, Climate Data Operators (CDO), and Python (https://www.python.org/) for producing the results.

**Data availability Statement:**

Observational and Reanalysis Data was derived from the following sources:

ERA5 (https://www.ecmwf.int/en/forecasts/dataset/ecmwf-reanalysis-v5); GPCP (https://psl.noaa.gov/data/gridded/data.gpcp.html) and TRMM (https://gpm.nasa.gov/data/directory). Data from model simulations are uploaded at Mendeley Data cloud server and can be accessed freely (https://data.mendeley.com/datasets/gy4hfbm9x9/1) . The codes for analysis and visualization are freely available from the following sites Ferret (https://ferret.pmel.noaa.gov/Ferret/); GrADS (https://cola.gmu.edu/grads/); NCL (https://www.ncl.ucar.edu/); CDO (https://code.mpimet.mpg.de/projects/cdo) and Python (https://www.python.org/).

# References

Adler, R. F., Huffman, G. J., Chang, A., Ferraro, R., Xie, P. P., Janowiak, J., et al. (2003). The version-2 global precipitation climatology project (GPCP) monthly precipitation analysis (1979-present). *Journal of Hydrometeorology*, *4*(6), 1147–1167. https://doi.org/10.1175/1525-7541(2003)004<1147:TVGPCP>2.0.CO;2

Baker, M. B. (1997). Cloud microphysics and climate. *Science*, *276*(5315), 1072–1078. https://doi.org/10.1126/science.276.5315.1072

Bhowmik, M., Hazra, A., Srivastava, A., Mudiar, D., Chaudhari, H. S., Rao, S. A., & Wang, L.-P. (2024). Improved Indian Summer Monsoon rainfall simulation: the significance of reassessing the autoconversion parameterization in coupled climate model. *Climate Dynamics*. https://doi.org/10.1007/s00382-024-07243-w

Boyle, J. S., Klein, S. A., Lucas, D. D., Ma, H. Y., Tannahill, J., & Xie, S. (2015). The parametric sensitivity of CAM5's MJO. *Journal of Geophysical Research*. https://doi.org/10.1002/2014JD022507

Chaudhari, H. S., Pokhrel, S., Kulkarni, A., Hazra, A., & Saha, S. K. (2016). Clouds–SST relationship and interannual variability modes of Indian summer monsoon in the context of clouds and SSTs: observational and modelling aspects. *International Journal of Climatology*, *36*(15), 4723–4740. https://doi.org/10.1002/joc.4664

Choudhury, B. A., Rajesh, P. V., Zahan, Y., & Goswami, B. N. (2022). Evolution of the Indian summer monsoon rainfall simulations from CMIP3 to CMIP6 models. *Climate Dynamics*, *58*(9–10), 2637–2662. https://doi.org/10.1007/s00382-021-06023-0

Dai, A. (2006). Precipitation characteristics in eighteen coupled climate models. *Journal of Climate*. https://doi.org/10.1175/JCLI3884.1

Dutta, U., Chaudhari, H. S., Hazra, A., Pokhrel, S., Saha, S. K., & Veeranjaneyulu, C. (2020). Role of convective and microphysical processes on the simulation of monsoon intraseasonal oscillation. *Climate Dynamics*, *55*(9–10), 2377–2403. https://doi.org/10.1007/s00382-020-05387-z

Dutta, U., Hazra, A., Chaudhari, H. S., Saha, S. K., Pokhrel, S., Shiu, C. J., & Chen, J. P. (2021). Role of Microphysics and Convective Autoconversion for the Better Simulation of Tropical Intraseasonal Oscillations (MISO and MJO). *Journal of Advances in Modeling Earth Systems*, *13*(10). https://doi.org/10.1029/2021MS002540

Dutta, U., Hazra, A., Chaudhari, H. S., Saha, S. K., Pokhrel, S., & Verma, U. (2022). Unraveling the global teleconnections of Indian summer monsoon clouds: expedition from CMIP5 to CMIP6. *Global and Planetary Change*, *215*, 103873. https://doi.org/10.1016/j.gloplacha.2022.103873

Dutta, U., Bhowmik, M., Hazra, A., Shiu, C.-J., & Chen, J.-P. (2024). Evaluation of the impact of the tropical oscillations on the Indian summer monsoon in the global climate models. *Theoretical and Applied Climatology*, *155*(9), 9007–9027. https://doi.org/10.1007/s00704-024-05160-w

Eyring, V., Bony, S., Meehl, G. A., Senior, C. A., Stevens, B., Stouffer, R. J., & Taylor, K. E. (2016). Overview of the Coupled Model Intercomparison Project Phase 6 (CMIP6) experimental design and organization. *Geoscientific Model Development*, *9*(5), 1937–1958. https://doi.org/10.5194/gmd-9-1937-2016

Field, P. R., & Heymsfield, A. J. (2015). Importance of snow to global precipitation. *Geophysical Research Letters*, *42*(21), 9512–9520. https://doi.org/10.1002/2015GL065497

Ganai, M., Krishna, R. P. M., Tirkey, S., Mukhopadhyay, P., Mahakur, M., & Han, J. Y. (2019). The Impact of Modified Fractional Cloud Condensate to Precipitation Conversion Parameter in Revised Simplified Arakawa-Schubert Convection Parameterization Scheme on the Simulation of Indian Summer Monsoon and Its Forecast Application on an Extreme Rainfa. *Journal of Geophysical Research: Atmospheres*. https://doi.org/10.1029/2019JD030278

Goswami, B. N. (1987). A mechanism for the west-north-west movement of monsoon depressions. *Nature*, *326*(6111), 376–378. https://doi.org/10.1038/326376a0

Goswami, B. N., Ajayamohan, R. S., Xavier, P. K., & Sengupta, D. (2003). Clustering of synoptic activity by Indian summer monsoon intraseasonal oscillations. *Geophysical Research Letters*, *30*(8). https://doi.org/10.1029/2002GL016734

Griffies, S. M., Gnanadesikan, A., Dixon, K. W., Dunne, J. P., Gerdes, R., Harrison, M. J., et al. (2005). Formulation of an ocean model for global climate simulations. *Ocean Science*. https://doi.org/10.5194/os-1-45-2005

Gusain, A., Ghosh, S., & Karmakar, S. (2020). Added value of CMIP6 over CMIP5 models in simulating Indian summer monsoon rainfall. *Atmospheric Research*, *232*, 104680. https://doi.org/10.1016/j.atmosres.2019.104680

Han, J., & Pan, H. L. (2011). Revision of convection and vertical diffusion schemes in the NCEP Global Forecast System. *Weather and Forecasting*, *26*(4), 520–533. https://doi.org/10.1175/WAF-D-10-05038.1

Han, J. Y., Hong, S. Y., Lim, K. S. S., & Han, J. (2016). Sensitivity of a cumulus parameterization scheme to precipitation production representation and its impact on a heavy rain event over Korea. *Monthly Weather Review*. https://doi.org/10.1175/MWR-D-15-0255.1

Hazra, A., Chaudhari, H. S., Saha, S. K., & Pokhrel, S. (2017). Effect of cloud microphysics on Indian summer monsoon precipitating clouds: A coupled climate modeling study. *Journal of Geophysical Research*, *122*(7), 3786–3805. https://doi.org/10.1002/2016JD026106

Hazra, A., Chaudhari, H. S., Saha, S. K., Pokhrel, S., & Goswami, B. N. (2017). Progress Towards Achieving the Challenge of Indian Summer Monsoon Climate Simulation in a Coupled Ocean-Atmosphere Model. *Journal of Advances in Modeling Earth Systems*, *9*(6), 2268–2290. https://doi.org/10.1002/2017MS000966

Hazra, A., Chaudhari, H. S., Saha, S. K., Pokhrel, S., Dutta, U., & Goswami, B. N. (2020). Role of cloud microphysics in improved simulation of the Asian monsoon quasi-biweekly mode (QBM). *Climate Dynamics*, *54*(1–2), 599–614. https://doi.org/10.1007/s00382-019-05015-5

Hazra, A., Dutta, U., Chaudhari, H. S., Pokhrel, S., & Konwar, M. (2023). Role of mean, variability and teleconnection of clouds behind Indian summer monsoon rainfall. *International Journal of Climatology*, *43*(9), 4099–4118. https://doi.org/10.1002/joc.8076

Hersbach, H., Bell, B., Berrisford, P., Hirahara, S., Horányi, A., Muñoz-Sabater, J., et al. (2020). The ERA5 global reanalysis. *Quarterly Journal of the Royal Meteorological Society*, *146*(730), 1999–2049. https://doi.org/10.1002/qj.3803

Hong, S. Y., Kwon, Y. C., Kim, T. H., Esther Kim, J. E., Choi, S. J., Kwon, I. H., et al. (2018). The Korean Integrated Model (KIM) System for Global Weather Forecasting. *Asia-Pacific Journal of Atmospheric Sciences*. https://doi.org/10.1007/s13143-018-0028-9

Hong, Y., Liu, G., & Li, J. L. F. (2016). Assessing the radiative effects of global ice clouds based on CloudSat and CALIPSO measurements. *Journal of Climate*, *29*(21), 7651–7674. https://doi.org/10.1175/JCLI-D-15-0799.1

Huffman, G. J., Adler, R. F., Bolvin, D. T., Gu, G., Nelkin, E. J., Bowman, K. P., et al. (2007). The TRMM Multisatellite Precipitation Analysis (TMPA): Quasi-global, multiyear, combined-sensor precipitation estimates at fine scales. *Journal of Hydrometeorology*. https://doi.org/10.1175/JHM560.1

Jain, S., Salunke, P., Mishra, S. K., & Sahany, S. (2019). Performance of CMIP5 models in the simulation of Indian summer monsoon. *Theoretical and Applied Climatology*, *137*(1–2), 1429–1447. https://doi.org/10.1007/s00704-018-2674-3

Jiang, X., & Ting, M. (2019). Intraseasonal variability of rainfall and its effect on interannual variability across the Indian subcontinent and the tibetan plateau. *Journal of Climate*, *32*(8), 2227–2245. https://doi.org/10.1175/JCLI-D-18-0319.1

Kumar, S., Hazra, A., & Goswami, B. N. (2014). Role of interaction between dynamics, thermodynamics and cloud microphysics on summer monsoon precipitating clouds over the Myanmar Coast and the Western Ghats. *Climate Dynamics*, *43*(3–4), 911–924. https://doi.org/10.1007/s00382-013-1909-3

Liebmann, B., & Smith, C. A. (1996). A description of a complete (interpoled) outgoing longwave radiation dataset. *Buletim of the American Meteorological Society*, *77*(6), 1275–1277.

Liu, Y., & Daum, P. H. (2004). Parameterization of the autoconversion process. Part I: Analytical formulation of the Kessler-type parameterization. *Journal of the Atmospheric Sciences*, *61*(13), 1539–1548. https://doi.org/10.1175/1520-0469(2004)061<1539:POTAPI>2.0.CO;2

Liu, Y., Daum, P. H., & McGraw, R. (2004). An analytical expression for predicting the critical radius in the autoconversion parameterization. *Geophysical Research Letters*. https://doi.org/10.1029/2003gl019117

Liu, Y., Daum, P. H., McGraw, R., & Miller, M. (2006). Generalized threshold function accounting for effect of relative dispersion on threshold behavior of autoconversion process. *Geophysical Research Letters*. https://doi.org/10.1029/2005GL025500

Maloney, E. D., & Hartmann, D. L. (2001). The Madden-Julian oscillation, barotropic dynamics, and North Pacific tropical cyclone formation. Part I: Observations. *Journal of the Atmospheric Sciences*. https://doi.org/10.1175/1520-0469(2001)058<2545:TMJOBD>2.0.CO;2

Meehl, G. A., Covey, C., Delworth, T., Latif, M., McAvaney, B., Mitchell, J. F. B., et al. (2007). The WCRP CMIP3 multimodel dataset: A new era in climatic change research. *Bulletin of the American Meteorological Society*, *88*(9), 1383–1394. https://doi.org/10.1175/BAMS-88-9-1383

Michalakes, J. (2020). HPC for Weather Forecasting. In *Modeling and Simulation in Science, Engineering and Technology* (pp. 297–323). https://doi.org/10.1007/978-3-030-43736-7_10

Michibata, T., & Takemura, T. (2015). Evaluation of autoconversion schemes in a single model framework with satellite observations. *Journal of Geophysical Research*. https://doi.org/10.1002/2015JD023818

Murakami, T. (1980). Temporal variations of satellite-observed outgoing longwave radiation over the winter monsoon region. Part II: short-period (4- 6 day) oscillations. *Monthly Weather Review*, *118*(4), 427–444. https://doi.org/10.1175/1520-0493(1980)108<0427:tvosoo>2.0.co;2

Pathak, R., Sahany, S., Mishra, S. K., & Dash, S. K. (2019). Precipitation Biases in CMIP5 Models over the South Asian Region. *Scientific Reports*. https://doi.org/10.1038/s41598-019-45907-4

Rajeevan, M., Gadgil, S., & Bhate, J. (2010). Active and break spells of the indian summer monsoon. *Journal of Earth System Science*, *119*(3), 229–247. https://doi.org/10.1007/s12040-010-0019-4

Rajeevan, M., Unnikrishnan, C. K., & Preethi, B. (2012). Evaluation of the ENSEMBLES multi-model seasonal forecasts of Indian summer monsoon variability. *Climate Dynamics*, *38*(11–12), 2257–2274. https://doi.org/10.1007/s00382-011-1061-x

Ramu, D. A., Sabeerali, C. T., Chattopadhyay, R., Rao, D. N., George, G., Dhakate, A. R., et al. (2016). Indian summer monsoon rainfall simulation and prediction skill in the CFSv2 coupled model: Impact of atmospheric horizontal resolution. *Journal of Geophysical Research*, *121*(5), 2205–2221. https://doi.org/10.1002/2015JD024629

Rao, S. A., Goswami, B. N., Sahai, A. K., Rajagopal, E. N., Mukhopadhyay, P., Rajeevan, M., et al. (2019). Monsoon mission a targeted activity to improve monsoon prediction across scales. *Bulletin of the American Meteorological Society*. https://doi.org/10.1175/BAMS-D-17-0330.1

Rotstayn, L. D. (2000). On the "tuning" of autoconversion parameterizations in climate models. *Journal of Geophysical Research Atmospheres*. https://doi.org/10.1029/2000JD900129

Saha, S., Moorthi, S., Pan, H. L., Wu, X., Wang, J., Nadiga, S., et al. (2010). The NCEP climate forecast system reanalysis. *Bulletin of the American Meteorological Society*. https://doi.org/10.1175/2010BAMS3001.1

Saha, S., Moorthi, S., Wu, X., Wang, J., Nadiga, S., Tripp, P., et al. (2014). The NCEP climate forecast system version 2. *Journal of Climate*. https://doi.org/10.1175/JCLI-D-12-00823.1

Saha, S. B., Roy, S. Sen, Roy Bhowmik, S. K., & Kundu, P. K. (2014). Unutarsezonska varijabilnost naoblake nad Indijskim potkontinentom tijekom monsunske sezone na temelju mjerenja oborine radarom TRMM. *Geofizika*, *31*(1), 29–53. https://doi.org/10.15233/gfz.2014.31.2

Saha, S. K., Hazra, A., Pokhrel, S., Chaudhari, H. S., Sujith, K., Rai, A., et al. (2019). Unraveling the Mystery of Indian Summer Monsoon Prediction: Improved Estimate of Predictability Limit. *Journal of Geophysical Research: Atmospheres*, *124*(4), 1962–1974. https://doi.org/10.1029/2018JD030082

Saha, S. K., Hazra, A., Pokhrel, S., Chaudhari, H. S., Rai, A., Sujith, K., et al. (2020). Reply to Comment by E. T. Swenson, D. Das, and J. Shukla on "Unraveling the Mystery of Indian Summer Monsoon Prediction: Improved Estimate of Predictability Limit." *Journal of Geophysical Research: Atmospheres*. https://doi.org/10.1029/2020JD033242

Song, X., & Zhang, G. J. (2011). Microphysics parameterization for convective clouds in a global climate model: Description and single-column model tests. *Journal of Geophysical Research Atmospheres*. https://doi.org/10.1029/2010JD014833

Sperber, K. R., Annamalai, H., Kang, I. S., Kitoh, A., Moise, A., Turner, A., et al. (2013). The Asian summer monsoon: An intercomparison of CMIP5 vs. CMIP3 simulations of the late 20th century. *Climate Dynamics*, *41*(9–10), 2711–2744. https://doi.org/10.1007/s00382-012-1607-6

Suhas, E., Neena, J. M., & Goswami, B. N. (2013). An Indian monsoon intraseasonal oscillations (MISO) index for real time monitoring and forecast verification. *Climate Dynamics*, *40*(11–12), 2605–2616. https://doi.org/10.1007/s00382-012-1462-5

Sundqvist, H., Berge, E., & Kristjansson, J. E. (1989). Condensation and cloud parameterization studies with a mesoscale numerical weather prediction model. *Monthly Weather Review*. https://doi.org/10.1175/1520-0493(1989)117<1641:CACPSW>2.0.CO;2

Taylor, K. E., Stouffer, R. J., & Meehl, G. A. (2012). An overview of CMIP5 and the experiment design. *Bulletin of the American Meteorological Society*, *93*(4), 485–498. https://doi.org/10.1175/BAMS-D-11-00094.1

Weber, T., & Quaas, J. (2012). Incorporating the subgrid-scale variability of clouds in the autoconversion parameterization using a PDF-scheme. *Journal of Advances in Modeling Earth Systems*. https://doi.org/10.1029/2012MS000156

Wheeler, M., & Kiladis, G. N. (1999). Convectively Coupled Equatorial Waves: Analysis of Clouds and Temperature in the Wavenumber-Frequency Domain. *Journal of the Atmospheric Sciences*, *56*(3), 374–399. https://doi.org/10.1175/1520-0469(1999)056<0374:CCEWAO>2.0.CO;2

Zhang, J., Lohmann, U., & Lin, B. (2002). A new statistically based autoconversion rate parameterization for use in large-scale models. *Journal of Geophysical Research Atmospheres*. https://doi.org/10.1029/2001JD001484

Zhao, Q., & Carr, F. H. (1997). A prognostic cloud scheme for operational NWP models. *Monthly Weather Review*. https://doi.org/10.1175/1520-0493(1997)125<1931:APCSFO>2.0.CO;2

Zhu, Y., & Yang, S. (2021). Interdecadal and interannual evolution characteristics of the global surface precipitation anomaly shown by CMIP5 and CMIP6 models. *International Journal of Climatology*, *41*(S1), E1100–E1118. https://doi.org/10.1002/joc.6756